\documentstyle[PASJadd,psfig]{PASJ95}
\draft

\newcommand{\vol}{51}
\newcommand{\no}{0}
\newcommand{\lett}{}
\newcommand{\spage}{1}
\newcommand{\rdate}{1998 October 8}
\newcommand{\adate}{1999 February 23}

\begin{document}
\setcounter{page}{1}


\title{ASCA Observation of the Low-Luminosity Seyfert 1.5 Galaxy NGC 5033}

\author{Yuichi {\sc Terashima}\thanks{%
Present address: NASA Goddard Space Flight Center, Code 662, Greenbelt, 
MD 20771, USA.},
Hideyo {\sc Kunieda}, and Kazutami {\sc Misaki}\\
{\it Department of Physics, Nagoya University, Chikusa-ku, Nagoya 464-8602}\\
}



\abst{ We present the results of an ASCA observation of the
low-luminosity Seyfert 1.5 galaxy NGC 5033. A point-like X-ray source
with a luminosity of $2.3\times10^{41}$ erg s$^{-1}$ in the 2--10 keV
band (at 18.7 Mpc; Tully 1988, AAA045.002.054) was detected at the
nucleus. The X-ray light curve shows variability on a timescale of
$\sim10^4$ s with an amplitude of $\sim$20\%. The X-ray continuum is
represented by a weakly absorbed ({$N_{\rm H}$}
$\approx9\times10^{20}$ {cm$^{-2}$}) power-law with a photon index of
$1.72\pm0.04$, which is quite similar to Seyfert 1 galaxies with
higher luminosities. A Fe K$\alpha$ emission line is detected at
$6.40^{+0.08}_{-0.06}$~keV (redshift corrected) and the equivalent
width is $290\pm100$ eV. The line width is unresolved. The narrower
line width and larger equivalent width compared to Seyfert 1s imply
that fluorescent Fe K$\alpha$ emission from matter further out from
the center than the accretion disk significantly contributes to the
observed Fe K$\alpha$ line. We suggest that fluorescent Fe K$\alpha$
emission from the putative torus contributes to the observed Fe
K$\alpha$ line.}


\kword{Galaxies: active --- Galaxies: individual (NGC 5033) --- 
Galaxies: Seyfert --- X-Rays: spectra}



\maketitle
\thispagestyle{headings}

\section{Introduction}

Extensive studies of nearby galactic nuclei established that many
nearby galaxies harbor low-luminosity active galactic nuclei (LLAGNs)
(Ho 1999; Ho et al. 1997a, b). In the X-ray regime, their luminosities
are typically $10^{40}-10^{41}$ erg s$^{-1}$ in the energy range of
2--10~keV. X-ray observations of such AGNs provide us with a good
opportunity to study the accretion and emission mechanisms of AGNs
under extremely low-luminosity conditions. Owing to the high
sensitivity and imaging capability at energies above 2 keV of the ASCA
satellite (Tanaka et al. 1994), several LLAGNs have been detected
(Makishima et al. 1994; Ptak et al. 1996; Ishisaki et al. 1996;
Iyomoto et al. 1996, 1997, 1998; Serlemitsos et al. 1996; Nicholson et
al. 1998; Terashima et al. 1998a, b). ASCA observations have revealed
that the X-ray continuum shape in LLAGNs is quite similar to that of
luminous Seyfert galaxies in spite of their low luminosity. On the
other hand, the LLAGNs M81 and NGC 4579 show higher center energies of
the Fe K$\alpha$ emission, which are consistent with He-like Fe
(Ishisaki et al. 1996; Serlemitsos et al. 1996; Terashima et
al. 1998b), while no Fe K$\alpha$ emission is detected from NGC 1097
(Iyomoto et al. 1996), NGC 3065, and NGC 4203 (Iyomoto et
al. 1998). These Fe K$\alpha$ emission-line properties are different
from that of Seyfert galaxies, in which broad Fe K$\alpha$ emission is
generally detected at $\sim$ 6.4~keV (e.g. Nandra et
al. 1997b). Furthermore, LLAGNs do not show any rapid and large-amplitude
variability, in contrast to Seyfert galaxies (Ishisaki et al. 1996;
Iyomoto et al. 1996, 1998; Terashima et al. 1998b; Ptak et al. 1998;
Awaki et al. in preparation). In regular Seyfert 1 galaxies, lower
luminosity objects tend to show variability with large amplitude on
short timescales (Nandra et al. 1997a; Mushotzky et al. 1993). Such
differences between LLAGNs and Seyfert 1s have been attributed to the
difference in the structure of the accretion disks (Ptak et al. 1998)
or the central black-hole mass (Awaki et al. in preparation).  Further
X-ray observations of LLAGNs are of great importance to investigate
these differences between LLAGNs and `classical' AGNs in order to
understand the underlying physics in LLAGNs,

NGC 5033 is a nearby (18.7 Mpc, $z$ = 0.00292; Tully 1988; we use a
value of $H_0$ = 75 km s$^{-1}$ Mpc$^{-1}$ for the Hubble constant)
SBc galaxy with a Seyfert nucleus. The H$\alpha$ and H$\beta$ emission
lines have strong, broad components, leading to the classification of
NGC 5033 as a Seyfert 1.5 galaxy. The broad H$\alpha$ emission line is
variable (Ho et al. 1997a, b; Koratkar et al. 1995; Filippenko,
Sargent 1985; Stauffer 1982). The narrow H$\alpha$ luminosity is only
$\log L({\rm H}\alpha)$ = 39.32 (here $L$ in erg s$^{-1}$), which is
smaller than the well-studied low-luminosity Seyfert galaxy NGC 4051.

X-ray observations of NGC 5033 have been performed with various
instruments (Polletta et al. 1996 and references therein). In the soft
X-ray band, Einstein IPC detected an X-ray source at the nucleus
(Halpern, Steiner 1983), and the X-ray emission is unresolved in a
ROSAT HRI image (Koratkar et al. 1995). Although there are a few
hard X-ray observations above 2 keV, no spectral information has been
obtained so far. For example, EXOSAT detected X-ray emission
with a flux of $4.7\times10^{-12}$ erg s$^{-1}$ cm$^{-2}$ in the
2--10~keV band, but the X-ray spectral shape is not constrained because
of limited photon statistics and background systematics (Turner,
Pounds 1989).

In this paper, we report on the first measurements of an X-ray
spectrum and the detection of Fe K$\alpha$ emission.


\section{Observation}

We observed NGC 5033 with ASCA on 1995 December 14. A detailed
description of the ASCA instruments can be found in Serlemitsos et al. 
(1995), Ohashi et al. (1996), Makishima et al. (1996), Burke et
al. (1994), and Yamashita et al. (1997). The Solid-state Imaging
Spectrometers (SIS 0 and SIS 1) were operated in 1 CCD FAINT mode. The
Gas Imaging Spectrometers (GIS 2 and GIS 3) were operated in the PH
normal mode. We obtained net exposure times of 36 ks and 39 ks for SIS
and GIS, respectively, after standard data screening.  X-ray spectra
and light curves were extracted from circular regions centered on the
NGC 5033 nucleus with a radius of 4' for SIS and 6' for GIS. The
spectra from SIS 0 and SIS 1 were combined after gain corrections. The
spectra from GIS 2 and GIS 3 were also combined. Background data were
accumulated from a source-free region in the same field. The mean
count rates for SIS and GIS were 0.16 c s$^{-1}$ and 0.11 c s$^{-1}$
per detector, respectively, after background subtraction.

\section{Results}



An X-ray source is clearly detected at the position of the NGC 5033
nucleus within the accuracy of the position determination. Significant
X-ray variability is seen in all instruments. The combined X-ray
light curve of the SIS (0.5--10 keV) + GIS (0.7--10 keV) data with a
bin size of 5760 s (one orbit of the ASCA satellite) is shown
in figure 1. Intensity variations of $\sim20\%$ on timescales of $\sim$
$10^4$~s are clearly seen. A constant model fit to this light curve
yields $\chi^2$=106 for 14 degrees of freedom.  We calculated a
normalized excess variance, as defined in Nandra et al. (1997a), using
a light curve obtained with SIS 0+1 in the 0.5--10 keV band with a bin
size of 128 s, and obtained $\sigma^2_{\rm rms} =
(1.7\pm0.7)\times10^{-3}$.



The X-ray spectra obtained with SIS and GIS are shown in figure 2. The
X-ray continuum is well represented by a single power-law modified by
photoelectric absorption, and an emission-line feature is seen at around
6.4~keV. Therefore, we fit the overall spectra with an absorbed
power-law continuum plus a Gaussian model. This model provides a good
fit to the data ($\chi^2_{\nu}$ = 0.93 for 188 dof), and the best-fit
parameters are summarized in table 1., where the line is assumed to
have zero width. The best-fit model is shown as a histogram in figure
2. The photon index is $1.72\pm0.04$ and the absorption
column density ($N_{\rm H}$ =$8.7\pm1.7\times10^{20}$~cm$^{-2}$) is
slightly higher than the galactic value of
$1.1\times10^{20}$~cm$^{-2}$ (Murphy et al. 1996) (hereafter the
quoted errors are at the 90\% confidence level for one interesting
parameter).  The X-ray flux in the 2--10~keV band is
$5.5\times10^{-12}$~erg s$^{-1}$ cm$^{-2}$, which corresponds to
$2.3\times10^{41}$~erg s$^{-1}$ at a distance of 18.7 Mpc. The error
for the flux is $\sim$10\%, which is dominated by calibration
uncertainties. The addition of a Gaussian line to a power-law continuum model
significantly improved the $\chi^2$ value by $\Delta\chi^2=-22.5$ for
two additional parameters (line center energy and
intensity). According to the $F$-test, this component is statistically
significant at 99.999\% confidence. The best-fit line center energy
and equivalent width are $6.38^{+0.08}_{-0.06}$~keV (observed frame)
and $290\pm100$~eV, respectively. This line centroid energy indicates
that the origin of the emission is fluorescence from cold iron. If the
line width is allowed to vary, the $\chi^2$ value improves by only
$\Delta\chi^2=-0.4$ and the line center, width, and equivalent
width become $6.41^{+0.14}_{-0.09}$~keV, $100^{+160}_{-100}$~eV, and
$320^{+110}_{-140}$~eV, respectively. Thus, the line is unresolved. The
upper limit of the line width, $\sigma\approx260$ eV, corresponds to
$\sigma\approx$12000 km s$^{-1}$.  The spectrum around the Fe
K$\alpha$ emission is shown in figure~3 as ratios of the data to the
best-fit continuum model. Confidence contours for the line energy
versus intensity and for the line width versus intensity are shown in
figures~4 and 5, respectively. The low energy (red) side of the
profile is steep, in contrast to the line profiles typical of Seyfert
1 galaxies, which have an extended red wing (e.g. Nandra et al. 1997b).
The Fe line profile indicates a hint of a wing at higher energies from
the peak. However, the addition of a second, Gaussian line at 6.7 keV
gives no significant improvement in the fit ($\Delta \chi^2 = -2.0$).




  Many Seyfert 1 galaxies show an Fe K$\alpha$ line profile skewed to
lower energies. This is interpreted in terms of the origin of the line
in the innermost part of an accretion disk (e.g. Tanaka et al. 1995;
Nandra et al. 1997b). We also apply the disk-line model (Fabian et al. 
1989) to the Fe K$\alpha$ emission from NGC 5033. Since the photon
statistics are limited, we fix the line center energy at 6.4~keV and
the inner radius of the line emitting region at 6$r_{\rm g}$, where
$r_{\rm g} = GM/c^2$ is the gravitational radius.  The line emissivity
is assumed to be proportional to $r^{-q}$ and $q$ is fixed at 2.5,
which is a typical value for Seyfert 1 galaxies (Nandra et al. 1997b).
In this model fit, the outer radius of the line emitting region
becomes greater than 1000$r_{\rm g}$, indicating that the Fe line is
mainly emitted from an extended region in which relativistic effects
are not significant and that the line width is relatively narrow. This
is consistent with the Gaussian-fit result described
above. Additionally, the observed line profile shows no significant
red asymmetry. We tentatively fixed the outer radius at 1000$r_{\rm
g}$ and obtained a slightly worse fit ($\chi^2$=179.4 for 188 dof) than
the Gaussian model.  The best-fit inclination angle and equivalent
width are $37^{\circ +13}_{~-11} $ and $490^{+220}_{-210}$ eV,
respectively.  The results of various model fits to the Fe K$\alpha$
line are summarized in table 2.

  A thermal bremsstrahlung plus Gaussian model was also tried, but
provided a worse $\chi^2$ ($\Delta \chi^2$ = +25.5) than the power-law
plus Gaussian fit, and systematic negative and positive residuals are
seen below 0.7~keV and above 7~keV, respectively. Furthermore, the
best-fit center energy of the Gaussian component, 6.40~keV (redshift
corrected), indicates that the emission line is due to fluorescence
from neutral or low ionization-state ($<$Fe {\sc XVI}) iron. We therefore
conclude that a thermal model is not appropriate for the NGC~5033
spectrum.


Finally, we searched for the variability of the Fe K line intensity
within the observation.  We extracted the X-ray counts using the
 first 40.7 ks and the last 46.1 ks of the observation yielding
spectra in a low- and high-flux state, respectively. These spectra
were fitted with a power-law plus Gaussian model, where the line was
assumed to have zero width. The fitting results are also summarized in
table 1. No significant Fe K$\alpha$ line variability was detected
between these two spectra. The X-ray fluxes in the 2--10 keV band are
$5.0\times10^{-12}$~erg s$^{-1}$ cm$^{-2}$ and $6.0\times10^{-12}$~erg
s$^{-1}$ cm$^{-2}$, for low- and high-flux state, respectively. The
statistical errors (including the uncertainties in the spectral fits)
of these X-ray fluxes are $\sim$ 4\%, which are valid for comparisons
between the datasets analyzed here. The uncertainties for the absolute
fluxes are $\sim10$\%.

\section{Discussion}

\subsection{X-Ray Variability}

We detected X-ray emission with a luminosity of $2.3\times10^{41}$~erg
s$^{-1}$ from the nucleus of NGC 5033. X-ray variability on timescales
of $\sim10^4$ s was detected in a one day observation, and the
observed variability amplitude was only $\sim$ 20\%. Since less
luminous Seyfert 1 galaxies tend to show large amplitude and rapid
variability compared to luminous Seyfert 1 and quasars (Nandra et
al. 1997a; Lawrence, Papadakis 1993), the observed variability
amplitude is smaller than that expected from this trend. The
normalized excess variance, $\sigma^2_{\rm rms} = 1.7\times10^{-3}$,
is about two orders of magnitude smaller than the relation between
$\sigma_{\rm rms}$ and the X-ray luminosities for the Seyfert 1
galaxies reported in Nandra et al. (1997a) and similar to other LLAGNs
(Ptak et al. 1998).

The small-amplitude variability could be explained if hard X-ray
emitting components besides an AGN, such as X-ray binaries and
starburst activity, contribute to the X-ray flux significantly. Then,
the AGN component would be diluted, leading to a smaller observed
amplitude variability than that intrinsic to the AGN X-ray emission.
However, this possibility seems to be unlikely for the following
reasons. In normal spiral galaxies, X-ray emission mainly comes from
the superposition of discrete sources, such as low-mass X-ray
binaries, and the X-ray luminosity is roughly proportional to the
B-band luminosities $L_B$ (e.g. Fabbiano 1989). The large {$L_{\rm
X}$}/$L_B$ ratio ({$L_{\rm X}$}/$L_B$ = $1.5 \times10^{-3}$) for NGC
5033 compared to that of normal spiral galaxies (e.g. {$L_{\rm
X}$}/$L_B$ = $3.5\times10^{-5}$ for M31; Makishima et al. 1989)
combined with the poor fit to the thermal model characteristic of low
mass X-ray binaries suggests that the emission from X-ray binaries
does not contribute to the X-ray flux significantly. Also, there is no
evidence for the soft thermal X-ray emission of $kT\sim1$ keV which
accompanies starburst activity in the host galaxy (e.g. Tsuru et al.
1997; Moran, Lehnert 1997; Ptak et al. 1997). Additionally, the X-ray
continuum of NGC 5033 is well represented by a power-law, and thermal
bremsstrahlung emission does not fit the data, which appear to be
present in the X-ray spectrum of starburst galaxies (e.g. $kT\sim6$
keV for NGC 253, Persic et al. 1998). We conclude that the observed
small amplitude of the variation is an intrinsic property of the AGN in
NGC 5033.


NGC 5033 also shares the variability characteristics of previously
observed LLAGNs, and we confirmed that little or no variability is
quite common in low-luminosity AGNs with X-ray luminosities $< \sim
10^{41}$~erg s$^{-1}$, as suggested in Ptak et al. (1998) and Awaki et
al. (in preparation). We note that the low-luminosity Seyfert 1 NGC
4051 (e.g. {$L_{\rm X}$} $\sim 8\times10^{41}$~erg s$^{-1}$ at 17 Mpc;
Guainazzi et al. 1996), which is known to show rapid and
large-amplitude X-ray variability, is most likely to be a narrow-line
Seyfert 1 galaxy (Ho et al. 1997b; Osterbrock, Pogge 1985), which is a
subclass of Seyfert galaxies often showing rapid X-ray variability
(e.g. Boller et al. 1996). Thus, the negative correlation between the
amplitudes and luminosities is no longer seen at luminosities below
$\sim10^{41}$ erg s$^{-1}$.

\subsection{X-Ray Continuum}

The X-ray spectrum of NGC 5033 is well represented by the combination
of a power-law continuum and a Gaussian line. The continuum is fitted
by a power-law with a photon index of $\Gamma=1.72\pm0.04$, which is
very similar to Seyfert 1 galaxies with higher luminosities (Nandra et
al. 1997b). In Seyfert 1 galaxies, intrinsic photon indices are
steeper ($\Gamma\approx1.9$) than observed if the effect of Compton
reflection is taken into account. In the case of NGC 5033, the
intrinsic photon index cannot be constrained because of the limited
photon statistics and energy bandpass. The detection of strong,
fluorescent Fe K$\alpha$ emission implies that the reflection
component due to cold matter, which produces the Fe emission, is
present. It is thus likely that the intrinsic photon index in NGC 5033
is steeper than that observed, and is also consistent with higher
luminosity Seyfert 1 galaxies. Thus, we do not find any evidence for
a luminosity dependence of the X-ray continuum slope.

\subsection{Fe K$\alpha$ Emission}

We detected an Fe K$\alpha$ emission line at 6.4~keV which is due to
fluorescence in cold iron. The detection of Fe K$\alpha$ from several
LLAGNs has been reported. Although strong fluorescent Fe K$\alpha$ is
detected from LLAGNs with large intrinsic absorption ({{$N_{\rm
H}$}}$>10^{23}$ {cm$^{-2}$}) (Makishima et al. 1994; Terashima et
al. 1998a; Maiolino et al. 1998), this is the first LLAGN with small
intrinsic absorption to show an Fe K$\alpha$ emission line at 6.4 keV. 
Since other LLAGNs with small intrinsic absorption show no Fe
K$\alpha$ emission (NGC 1097; Iyomoto et al. 1997) or an Fe K$\alpha$
line at $\sim6.7$~keV (M81 Ishisaki et al. 1996; Serlemitsos et
al. 1996; NGC 4579 Terashima et al. 1998b), we conclude that a variety
of center energies and equivalent widths of Fe lines is observed in
LLAGNs.

An Fe K$\alpha$ line peaking at 6.4 keV is generally observed from
Seyfert 1 galaxies, and the line width is often observed to be broad
and skewed to lower energies (Nandra et al. 1997b). Such Fe K$\alpha$
emission from Seyfert 1 galaxies is interpreted as originating in the
innermost part of an accretion disk (e.g. Tanaka et al. 1995). The Fe
K$\alpha$ line in NGC 5033 is unresolved, and no clear signature of the
profile skewed to lower energies is seen.  Moreover, the equivalent
width, $290\pm100$~eV, is larger than the equivalent width of the narrow
core of the Fe lines in Seyfert 1 galaxies (100--150~eV) (Nandra et
al. 1997b). These results suggest that the narrow fluorescent Fe
K$\alpha$ line emission from beyond the accretion disk, possibly the
putative molecular torus assumed in the unified scheme, contributes to
the observed Fe emission. Note that the Fe line profiles of Seyfert 1
galaxies are well fitted with the disk-line model by Fabian et
al. (1989) and only a weak additional narrow component is required in
all but a small number of objects (Nandra et al. 1997b). The torus out
of the line of sight could be absent or subtend a small solid angle
viewed from the nucleus in Seyfert 1 galaxies.

Variability of the Fe K$\alpha$ line intensity has been detected in
several bright Seyfert 1 galaxies on timescales of less than one day
(Iwasawa et al. 1996; Yaqoob et al 1996; Nandra et al. 1997c). Such
variability also supports that Fe K$\alpha$ emission comes from the
inner part ($<$ one light day) of the accretion disk. On the other
hand, no significant intensity variability has been observed within
the one-day observation of NGC 5033, although the photon statistics
are limited. This is consistent with the picture that the Fe K$\alpha$
in NGC 5033 is produced mainly in the putative torus.

Thus, the relatively narrow and strong Fe K$\alpha$ line in NGC 5033
can be explained if a narrow Fe line from the torus significantly
contributes to the observed Fe emission. The optical emission line
widths of NGC 5033 classify it as an intermediate type (Seyfert
1.5). This intermediate classification might also be explained if we
are seeing the central engine from an intermediate inclination angle
and the broad line region is partly obscured. Then, the inner surface
of the torus can account for the fluorescent Fe emission. Thus, we
suggest that the obscuring torus assumed in the unified scheme of
Seyfert galaxies is present around the central engine of NGC 5033 in
contrast to Seyfert 1 galaxies, in which the torus contribution to Fe
K$\alpha$ emission is not important (Nandra et al. 1997a), possibly
because the torus is absent or subtends a small solid angle viewed from
the nucleus.


Future spectroscopic observations with higher energy resolution should
be able to decompose the broad (diskline) and narrow (outer part of
the accretion disk and/or putative torus) components.






\par
\vspace{1pc}\par

The authors thank T. Yaqoob for a critical reading of the manuscript,
and all of the ASCA team members who made this study possible. YT
and KM thank JSPS for support.

\section*{References}
\small

\re
Boller Th., Brandt W.N., Fink H. 1996, A\&A 305 53		

\re
Burke B.E., Mountain R.W., Daniels P.J., Dolat V.S. 1994, IEEE
Trans. NS-41, 375

\re
Fabbiano G. 1989, ARA\&A 27, 87		

\re
Fabian A.C., Rees M.J., Stella L., White N.E. 1989, MNRAS 238, 729	

\re
Filippenko A.V., Sargent W.L.W. 1985, ApJS 57, 503		

\re
Guainazzi M., Mihara T., Otani C., Matsuoka M. 1996, PASJ 48, 781 

\re
Halpern J.P., Steiner J.E. 1983, ApJ 269, L37

\re 
Ho L.C. 1999, in The 32nd COSPAR meeting, The AGN-Galaxy connection
(Advances in Space Research, Oxford), in press

\re
Ho L.C., Filippenko A.V., Sargent W.L.W. 1997a, ApJS 112, 315	

\re
Ho L.C., Filippenko A.V., Sargent W.L.W., Peng C.Y. 1997b, ApJS 112, 391 

\re
Ishisaki Y., Makishima K., Iyomoto N., Hayashida K., Inoue H., Mitsuda K., Tanaka Y., Uno S. et al. 1996, PASJ 48, 237			

\re
Iwasawa K., Fabian A.C., Reynolds C.S., Nandra K., Otani C., Inoue H., Hayashida K., Brandt W.N. et al. 1996, MNRAS 282, 1038		

\re
Iyomoto N., Makishima K., Fukazawa Y., Tashiro M., Ishisaki Y. 1997, PASJ 49, 425						

\re
Iyomoto, N. Makishima, K. Fukazawa, Y. Tashiro, M., Ishisaki Y. Nakai N., Taniguchi Y. 1996, PASJ 48, 231			

\re
Iyomoto N., Makishima K., Matsushita K., Fukazawa Y., Tashiro M., Ohashi T. 1998, ApJ 503, 168

\re
Koratkar A., Deustua S.E., Heckman T., Filippenko A.V., Ho L.C., Rao M. 1995, ApJ 440, 132

\re Lawrence A., Papadakis I. 1993, ApJ 414, L85

\re Maiolino R., Salvati M., Bassani L., Dadina M., Della Ceca R.,
Matt G., Risaliti G., Zamorani G. 1998, A\&A 338, 781

\re
Makishima K., Fujimoto R., Ishisaki Y., Kii T., Loewenstein M., Mushotzky R., Serlemitsos P., Sonobe T. et al. 1994, PASJ 46, L77	

\re
Makishima K., Ohashi T., Hayashida K., Inoue H., Koyama K., Takano S., Tanaka Y., Yoshida A. et al. 1989, PASJ 41, 697			

\re
Makishima K., Tashiro M., Ebisawa K., Ezawa H., Fukazawa Y., Gunji S., Hirayama M., Idesawa E. et al. 1996, PASJ 48, 171		

\re
Moran E.C., Lehnert M.D. 1997, ApJ 478, 172		

\re
Murphy E.M., Lockman F.J., Laor A., Elvis M. 1996, ApJS 105, 369

\re
Mushotzky R.F., Done, C., Pounds K.A. 1993, ARA\&A 31, 717

\re
Nandra K., George I.M., Mushotzky R.F., Turner T.J., Yaqoob T. 1997a, ApJ 476, 70							

\re
Nandra K., George I.M., Mushotzky R.F., Turner T.J., Yaqoob T 1997b, ApJ 477, 602							

\re
Nandra K., George I.M., Mushotzky R.F., Turner T.J., Yaqoob T 1997c, MNRAS 284, L7

\re
Nicholson K.L., Reichert G.A., Mason K.O., Puchnarewicz E.M., Ho
L.C., Shields J.C., Filippenko A.V. 1998, MNRAS 300, 893 	

\re
Ohashi T., Ebisawa K., Fukazawa Y., Hiyoshi K., Horii M., Ikebe Y., Ikeda H., Inoue H. et al. 1996, PASJ 48, 157					

\re
Osterbrock D.E., Pogge R.W. 1985, ApJ 297, 166		

\re 
Persic M., Mariani S., Cappi M., Bassani L., Danese L., Dean A.J.,
Di Cocco G., Franceschini A. et al. A\&A 339, L33	

\re
Polletta, M., Bassani, L., Malaguti, G., Palumbo, G.G.C., Caroli, E. 1996, ApJS 106, 399

\re
Ptak A., Serlemitsos P., Yaqoob T., Mushotzky R., Tsuru T. 1997, AJ 113, 1286								
\re
Ptak A., Yaqoob T., Mushotzky R., Serlemitsos P., Griffiths R.
1998, ApJ 501, L37 					

\re
Ptak A., Yaqoob T., Serlemitsos P.J., Kunieda H., Terashima Y. 1996, ApJ 459, 542							

\re
Serlemitsos P.J., Jalota L., Soong Y., Kunieda H., Tawara Y., Tsusaka Y., Suzuki H., Sakima Y. et al. 1995, PASJ 47, 105

\re Serlemitsos P.J., Ptak A.F., Yaqoob T. 1996, in The Physics of
LINERs in View of Recent Observations, ed. Eracleous M., Koratkar A.,
Leitherer C., Ho L.C. (ASP, San Francisco) p70

\re
Stauffer J.R. 1982, ApJ 262, 66

\re
Tanaka Y., Inoue H., Holt S.S. 1994, PASJ 46, L37

\re
Tanaka Y., Nandra K., Fabian A.C., Inoue H., Otani C., Dotani T., Hayashida K., Iwasawa K. et al 1995, Nature 375, 659 		

\re
Terashima Y., Kunieda H., Misaki K., Mushotzky R.F., Ptak A.F.,
Reichert G.A. 1998b, ApJ 503, 212		

\re
Terashima Y., Ptak A.F., Fujimoto R., Itoh M., Kunieda H., Makishima K. 
Serlemitsos P.J. 1998a, ApJ 496, 210		

\re
Tsuru T., Awaki H., Koyama K., Ptak A. 1997, PASJ 49, 619

\re
Tully R.B. 1988, Nearby Galaxies Catalog, (Cambridge Univ. Press: Cambridge)

\re
Turner T.J., Pounds K.A 1989, MNRAS 240, 833

\re
Yamashita A., Dotani T., Bautz M., Crew G., Ezuka H., Gendreau K., Kotani T., Mitsuda K. et al. 1997, IEEE Trans. NS-44, 847

\re Yaqoob T., Serlemitsos P.J., Turner T.J., George I.M., Nandra
K. 1996, ApJ 470, L27				

\clearpage

\begin{table}[t]
\begin{center}
Table~1. Results of spectral fits to the SIS and GIS spectra of NGC 5033.$^*$
\end{center}
\vspace{6pt}
\tabcolsep 3pt
\begin{tabular*}{\textwidth}{@{\hspace{\tabcolsep}
\extracolsep{\fill}}p{6pc}cccccccc} 
\hline \hline\\[-6pt]
	& {$N_{\rm H}$} 	& Photon index     & Energy & EW & $\chi^2$/dof\\
	& [$10^{20}$~{cm$^{-2}$}] &		& [keV] & [eV] & &\\ 
\hline
Total \dotfill	& $8.7\pm1.7$	& $1.72\pm0.04$ & $6.38^{+0.08}_{-0.06}$ & $290\pm100$	& 174.0/188\\ 
Low \dotfill 	& $9.0\pm1.7$	& $1.77\pm0.07$	& $6.30^{+0.13}_{-0.09}$ & $350^{+150}_{-160}$	& 167.5/166\\
High \dotfill	& $8.8\pm2.4$	& $1.72\pm0.06$	& $6.48\pm0.12$		& $300^{+170}_{-140}$	& 202.1/198\\
\hline
\end{tabular*}
\vspace{6pt}\par\noindent
* Line widths are assumed to have zero width. Line energies 
are in the observed frame.
\end{table}

\begin{table}[t]
\begin{center}
Table~2. Model fits to the Fe K line.$^*$
\end{center}
\vspace{6pt}
\tabcolsep 3pt
\begin{tabular*}{\textwidth}{@{\hspace{\tabcolsep}
\extracolsep{\fill}}p{6pc}ccccc} 
\hline \hline\\[-6pt]
Model	&	Energy [keV] & $\sigma$ [eV] or& EW [eV] 	& $\chi^2$/dof\\
	&		& inclination [deg]	&		&\\
\hline
(1) Narrow Gaussian line \dotfill	& $6.38^{+0.08}_{-0.06}$ & 0	& $290\pm100$	& 174.0/188\\
(2) Broad Gaussian line\dotfill	& $6.41^{+0.14}_{-0.09}$ & $100^{+160}_{-100}$	& $320^{+110}_{-140}$	& 173.6/187\\ 	
(3) Two narrow		& $6.36^{+0.07}_{-0.06}$ & 0	& $260^{+100}_{-120}$ & 172.0/187\\
 Gaussian lines \dotfill 	& 6.68	& 0		& $90^{+140}_{-90}$	& \\	
(4) Disk-line \dotfill	& 6.38	& $37^{\circ +13}_{~-11}$ & $490^{+220}_{-210}$ & 179.4/188\\
\hline
\end{tabular*}
\vspace{6pt}\par\noindent
* The values without errors are frozen parameters. 
Line energies are in the observed frame.
\end{table}

\clearpage

\onecolumn


\begin{figure}[htb]
\centerline{
\psfig{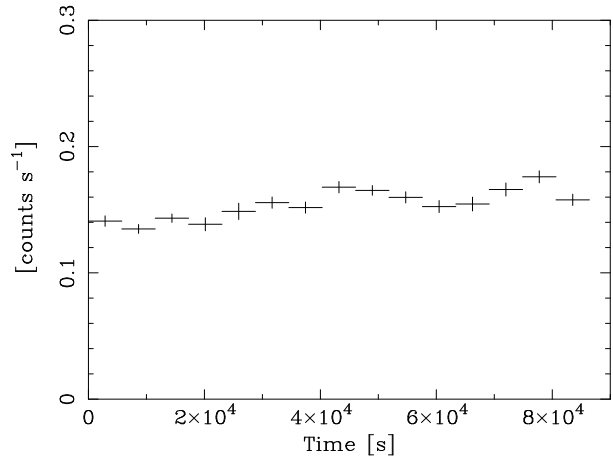}
}
\caption{X-ray light curve of NGC 5033. SIS (0.5--10~keV) and GIS 
(0.7--10~keV) data are combined. Bin size = 5760 s (one orbital
period of the ASCA satellite).}
        \label{fig:n5033_lc}
\end{figure}

\begin{figure}[htb]
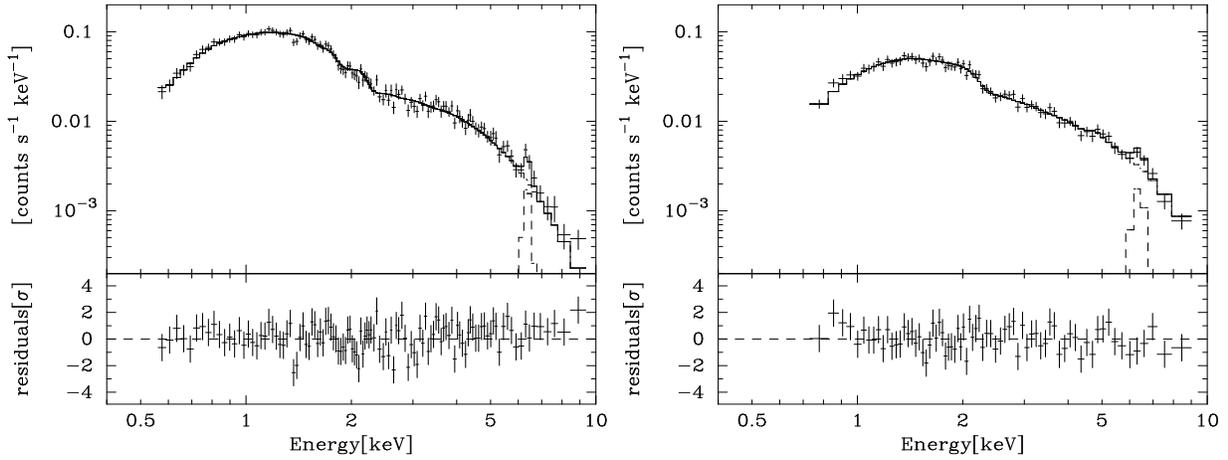

\centerline{
\psfig{file=fig2a.ps,width=8cm,height=6cm,angle=-90}
\psfig{file=fig2b.ps,width=8cm,height=6cm,angle=-90}
}
        \caption{X-ray spectra of NGC 5033. left: SIS, right: GIS. Though 
spectral fits are done simultaneously, the figures are shown separately for 
clarity. The histrograms indicate the best-fit power-law plus Gaussian model.}
        \label{fig:n5033_spec}
\end{figure}

\begin{figure}[htb]
\centerline{
\psfig{file=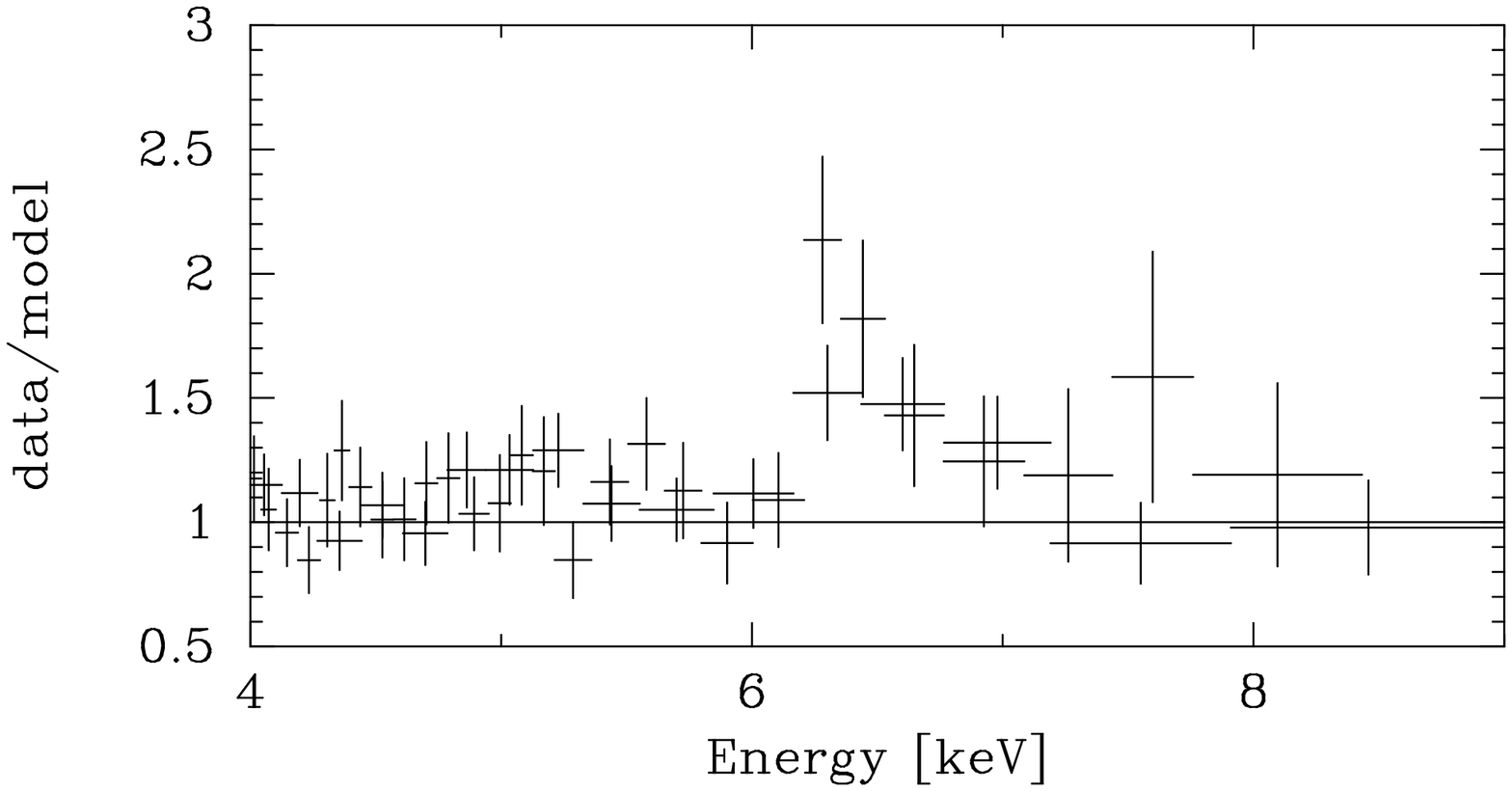,width=12cm,height=9cm,angle=-90}
}
        \caption{Data/model ratios around Fe K$\alpha$ emission in NGC 5033. 
The energy scale is not redshift corrected.}
        \label{fig:n5033_fe}

\centerline{
\psfig{file=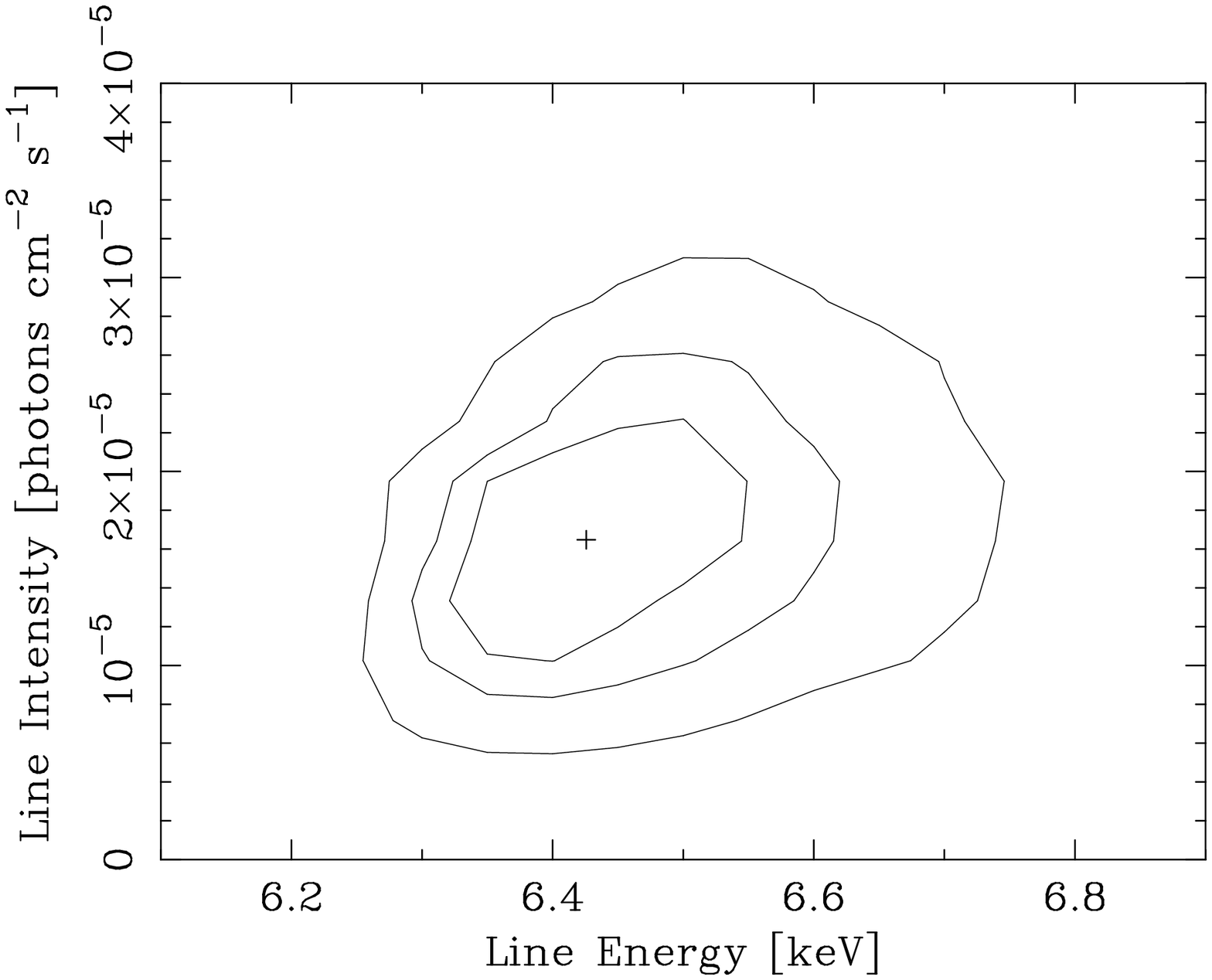,width=12cm,height=9cm,angle=-90}
}
\caption{Confidence contours for the line energy and
intensity. The line width is left to vary. The contours correspond to
68, 90, and 99\% confidence level for two interesting parameters
($\Delta \chi^2$ = 2.3, 4.6, and 9.2, respectively). The energy scale 
is not red shift corrected.} 
        \label{fig:n5033_fecont}
\end{figure}

\begin{figure}[hbt]
\centerline{
\psfig{file=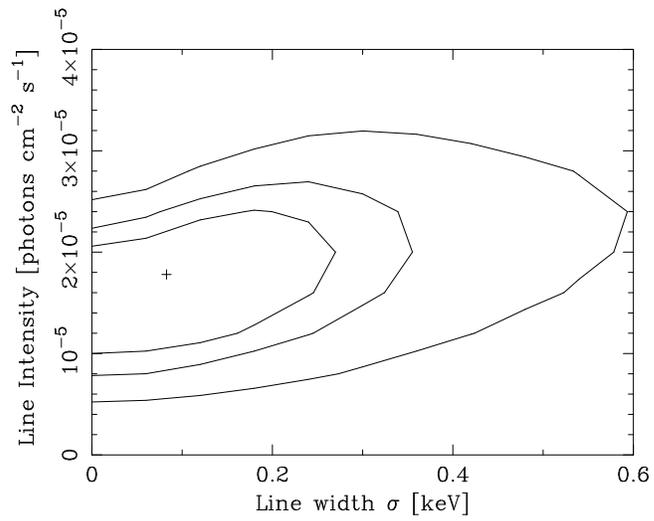,width=12cm,height=9cm,angle=-90}
}
\caption{Confidence contours for the line width and intensity. The contours correspond to 68, 90, and 99\% confidence level for two interesting parameters
($\Delta \chi^2$ = 2.3, 4.6, and 9.2, respectively).}
        \label{fig:n5033_fecont2}
\end{figure}

\end{document}